\documentclass[iop,apj,revtex4]{emulateapj}
\usepackage{graphicx}
\usepackage{amsmath,amssymb}

\usepackage{hyperref}

\begin{document}
\title{Plasma turbulence and kinetic instabilities at ion scales \\ in the expanding solar wind}

\author{Petr Hellinger\altaffilmark{1,2}}
\email{petr.hellinger@asu.cas.cz}
\author{Lorenzo Matteini\altaffilmark{3}}
\author{Simone Landi\altaffilmark{4,5}}
\author{Andrea Verdini\altaffilmark{4,6}}
\author{Luca Franci\altaffilmark{4,7}}
\author{Pavel~M.~Tr\'avn\'\i\v cek\altaffilmark{8,1,2}}

\altaffiltext{1}{Astronomical Institute, CAS,
Bocni II/1401,CZ-14100 Prague, Czech Republic}
\altaffiltext{2}{Institute of Atmospheric Physics, CAS,
Bocni II/1401, CZ-14100 Prague, Czech Republic}
\altaffiltext{3}{Department of Physics, Imperial College London, London SW7 2AZ, UK}
\altaffiltext{4}{Dipartimento di Fisica e Astronomia, Universit\`a degli Studi di Firenze Largo E. Fermi 2, I-50125 Firenze, Italy}
\altaffiltext{5}{INAF -- Osservatorio Astrofisico di Arcetri, Largo E. Fermi 5, I-50125 Firenze, Italy}
\altaffiltext{6}{Solar-Terrestrial Center of Excellence-SIDC, Royal Observatory of Belgium, Brussels, Belgium}
\altaffiltext{7}{INFN -- Sezione di Firenze, Via G. Sansone 1, I-50019 Sesto F.no (Firenze), Italy}
\altaffiltext{8}{Space Sciences Laboratory, University of California, 7 Gauss Way, Berkeley, CA 94720, USA}

\begin{abstract}
The relationship between a decaying strong turbulence and kinetic instabilities in a slowly expanding plasma
is investigated using
two-dimensional (2-D) hybrid expanding box simulations. 
We impose an initial ambient magnetic
field perpendicular to the simulation box, and we start with a spectrum of
large-scale, linearly-polarized, random-phase Alfv\'enic fluctuations which have energy equipartition between kinetic and magnetic
fluctuations and vanishing correlation between the two fields. A turbulent cascade rapidly develops, magnetic field fluctuations
 exhibit a power-law spectrum at large scales and
a steeper spectrum at ion scales.
The turbulent cascade leads to an overall anisotropic proton heating, protons
are heated in the perpendicular direction, and, initially, also in the parallel
direction. The imposed expansion leads to generation of
a large parallel proton temperature anisotropy which is at later stages partly reduced by turbulence. 
The turbulent heating is not sufficient to overcome the expansion-driven perpendicular cooling 
and the system eventually drives the oblique firehose instability
in a form of localized nonlinear wave packets which
 efficiently reduce the parallel temperature anisotropy.
This work demonstrates that kinetic instabilities may coexist
with strong plasma turbulence even in a constrained 2-D regime.
\end{abstract}
\pacs{?}
\maketitle

\section{Introduction}
Turbulence in magnetized
weakly collisional space and astrophysical plasmas is a ubiquitous nonlinear phenomenon 
that allows energy transfer from large to small scales, and, eventually, to plasma particles.
 Properties of plasma turbulence and its dynamics
remain an open challenging problem \citep{petral10,mave11}.
The solar wind constitutes a natural laboratory for 
plasma turbulence \citep{brca13,alexal13}, since it offers the opportunity of its detailed diagnostics.
Turbulence at large scales
can be described by the magnetohydrodynamic (MHD) approximation,
 accounting for the dominant nonlinear coupling and 
for the presence of the ambient magnetic field that introduces a preferred
direction \citep{boldal11}. 
Around 
particle characteristic scales 
 the plasma description has to be extended beyond MHD and, at these scales,
a transfer of the cascading energy to particles is expected.
The solar wind turbulence indeed likely energizes particles: 
radial profiles of proton temperatures indicate an important heating
which is often comparable to the estimated turbulent energy cascade rate \citep{macbal08,cranal09,hellal13}.
This energization proceeds through collisionless processes which may have a feedback on turbulence.
In the solar wind the problem is further complicated by a radial expansion 
 which induces an additional damping; 
turbulent fluctuations decrease due to the expansion as well as due to the turbulent decay.
The expansion thus slows down the turbulent
cascade \cite[cf.,][]{grapal93,dongal14}). Furthermore, the characteristic particle
scales change with radial distance affecting
possible particle energization mechanisms.

Understanding of the complex nonlinear properties of plasma turbulence on particle scales 
is facilitated via a numerical approach \citep{serval15,franal15b}.
Direct kinetic simulations of turbulence show that particles are indeed on average heated by the cascade 
\citep{paraal09,mava11,wual13,franal15b}, and, moreover, turbulence leads locally to
complex anisotropic and nongyrotropic distribution functions
\citep{valeal14,serval15}. Furthermore, expansion
naturally generates particle temperature anisotropies \citep{mattal12}. The anisotropic and nongyrotropic features
may be a source of free energy for kinetic instabilities.
In situ observations indicate existence of apparent bounds on the proton 
temperature anisotropies which are consistent with
theoretical kinetic linear predictions \citep{hellal06,hetr14}.
These linear predictions have, however, many limited assumptions \citep{mattal12,isenal13};
especially, they assume a homogeneous plasma which 
is at odds with the presence of turbulent fluctuations.
On the other hand, the observed bounds on the proton temperature anisotropy
(and other plasma parameters)
and enhanced magnetic fluctuations near these bounds
\citep{wickal13,lacoal14}
 indicate that these kinetic instabilities
are active even in presence of turbulence.

\section{Simulation results}
In this letter we directly test the relationship between
proton kinetic instabilities and plasma turbulence in the solar wind using
a hybrid expanding box model that allows
to study self-consistently physical processes at ion scales.
In the hybrid expanding box model
a constant solar wind radial velocity $v_{sw}$ is assumed.
The radial distance $R$
is then
$R = R_0 (1+ t/t_{e0})$
where $R_0$ is the initial position and $t_{e0}=R_0/v_{sw}$
 is the initial value of the characteristic expansion time $t_e=R/v_{sw}=t_{e0}(1+t/t_{e0})$.
Transverse scales (with respect to the radial direction)
 of a small portion of plasma, co-moving with the solar
wind velocity, increase 
$\propto R$.
The expanding box uses these co-moving coordinates,
approximating the spherical coordinates by the Cartesian ones \citep{liewal01,hetr05}.
The model uses the hybrid approximation where electrons are
considered as a massless, charge neutralizing fluid and
ions are described by a particle-in-cell model \citep{matt94}.
Here we use the two-dimensional (2-D) version of the code,
fields and moments are defined on a 2-D $x$--$y$ grid 
$2048 \times 2048$; periodic boundary conditions are assumed.
The spatial resolution is
$\Delta x=\Delta y=  0.25 d_{p0}$ where $d_{p0}=v_{A0}/\Omega_{p0}$ is
the initial proton inertial length ($v_{A0}$: the initial Alfv\'en velocity,
$\Omega_{p0}$: the initial proton gyrofrequency).
There are $1,024$ macroparticles per cell
for protons which are advanced
with a time step $\Delta t=0.05/\Omega_{p0}$
while the magnetic field 
is advanced with a smaller time step $\Delta t_B = \Delta t/10$.
The initial ambient magnetic field is directed along the radial, $z$ direction,
perpendicular to the simulation plane
$\boldsymbol{B}_{0}=(0,0,B_{0})$ and we impose a
continuous expansion in $x$ and $y$ directions.
Due to the expansion
the ambient density and the magnitude of the ambient magnetic field
decrease as $\bar{n}\propto \bar{B} \propto R^{-2}$ (the proton
inertial length $d_{p}$ increases $\propto R$,
the ratio between the transverse sizes and $d_{p}$ remains constant;
the proton gyrofrequency $\Omega_{p}$ decreases as $\propto R^{-2}$).
A small resistivity $\eta$ is used to avoid accumulation of
cascading energy at grid scales;  
initially we set $\eta = 10^{-3} \mu_0 v_{A0}^2/\Omega_{p0}$
($\mu_0$ being the magnetic permittivity of vacuum)
and $\eta$ is assumed to be 
$\propto \bar{n}$.
The simulation is initialized with an isotropic 2-D spectrum
of modes with random phases, linear Alfv\'en polarization ($\delta
\boldsymbol{B} \perp \boldsymbol{B}_0$), and vanishing correlation between
magnetic and velocity fluctuation. These modes are in the range 
 $0.02 \le k d_{p} \le 0.2$ and have a flat one-dimensional (1-D) power spectrum with rms fluctuations $=0.24 B_0$.
For noninteracting zero-frequency Alfv\'en waves the linear approximation predicts $\delta B_\perp\propto R^{-1}$ \citep{dongal14}.
Protons have initially the parallel proton beta  $\beta_{p\|} =  0.8$
and the parallel temperature anisotropy
 $A_{p}=T_{p\perp}/T_{p\|}=0.5$ as typical
proton parameters in the solar wind in the vicinity of 1~AU \citep{hellal06,marsal06}.
Electrons are assumed to be isotropic and isothermal with $\beta_{e} =   0.5$
at $t=0$.

\begin{figure}[htb]
\centerline{\includegraphics[width=8cm]{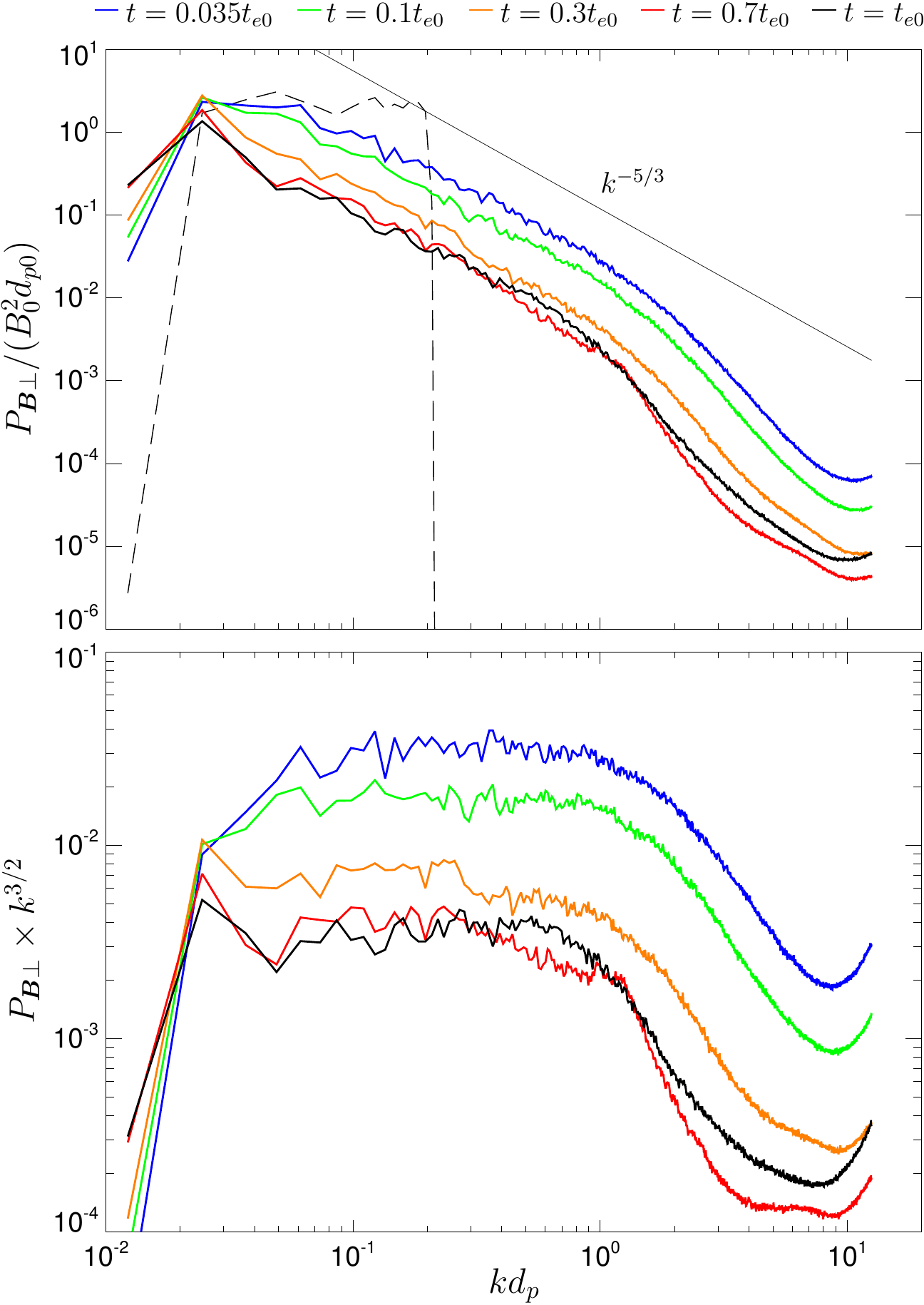}}
\caption{
(top) 1-D PSD $P_{\boldsymbol{B}\perp}$ of the fluctuating magnetic field
 $B_\perp$ perpendicular to the ambient magnetic field and (bottom)
$P_{\boldsymbol{B}\perp}$ compensated by $k^{3/2}$
 as functions of $k$ at different times.
The thin long dashed line shows the initial spectrum and
the thin solid line
shows a dependence $\propto k^{-5/3}$ for comparison.
\label{spec}}
\end{figure}

The initial random fluctuations rapidly
relax and a turbulent cascade develops.
Figure~\ref{spec} shows the evolution of the 1-D power spectral density (PSD) $P_{\boldsymbol{B}\perp}=P_{\boldsymbol{B}\perp}(k)$
of the magnetic field $\boldsymbol{B}_\perp$ perpendicular to $\boldsymbol{B}_0$.
On large scales, the initial flat spectrum evolves to a power law.
This large-scale power law remains clearly visible until $t\sim 0.7
t_{e0}$ although its slope slowly varies in time, passing from
about -3/2 to -5/3 (these estimated slopes are, however, quite
sensitive to the chosen range of wave vectors). The variation of
large-scale slopes ($k d_{p} \lesssim 1$) is likely connected with the decay of the large scale
fluctuations due to the cascade and the expansion as the inertial
range is likely quite narrow. This problem is beyond the scope of the
present letter and will be a subject of future work \cite[note that a similar steepening is also observed in
MHD expanding box simulations, cf.,][]{dongal14}; this letter is mainly focused on ion scales.

Around $k d_{p}\sim 1$ there is a smooth transition in $P_{\boldsymbol{B}\perp}$ separating 
a the large-scale power law slope and a steeper slope at sub-ion scales \citep{franal15a}.
The PSD amplitudes decay in time partly due to the expansion and
partly to the turbulent damping. Note that there are some indications that
the position of the transition shifts to smaller $k d_{p}$
with time/radial distance (compare the blue, green, orange, and red curves in Figure~\ref{spec}); a similar trend is observed for the proton
gyroradius since it 
increases only slightly faster than $d_{p}$.
At later times the fluctuating magnetic energy is enhanced at
ion scales around $k d_{p}\sim 0.4\div 1$ (compare the red and black curves in Figure~\ref{spec});
this indicates that some electromagnetic fluctuations are generated at later times
of the simulation.

\begin{figure}[htb]
\centerline{\includegraphics[width=8cm]{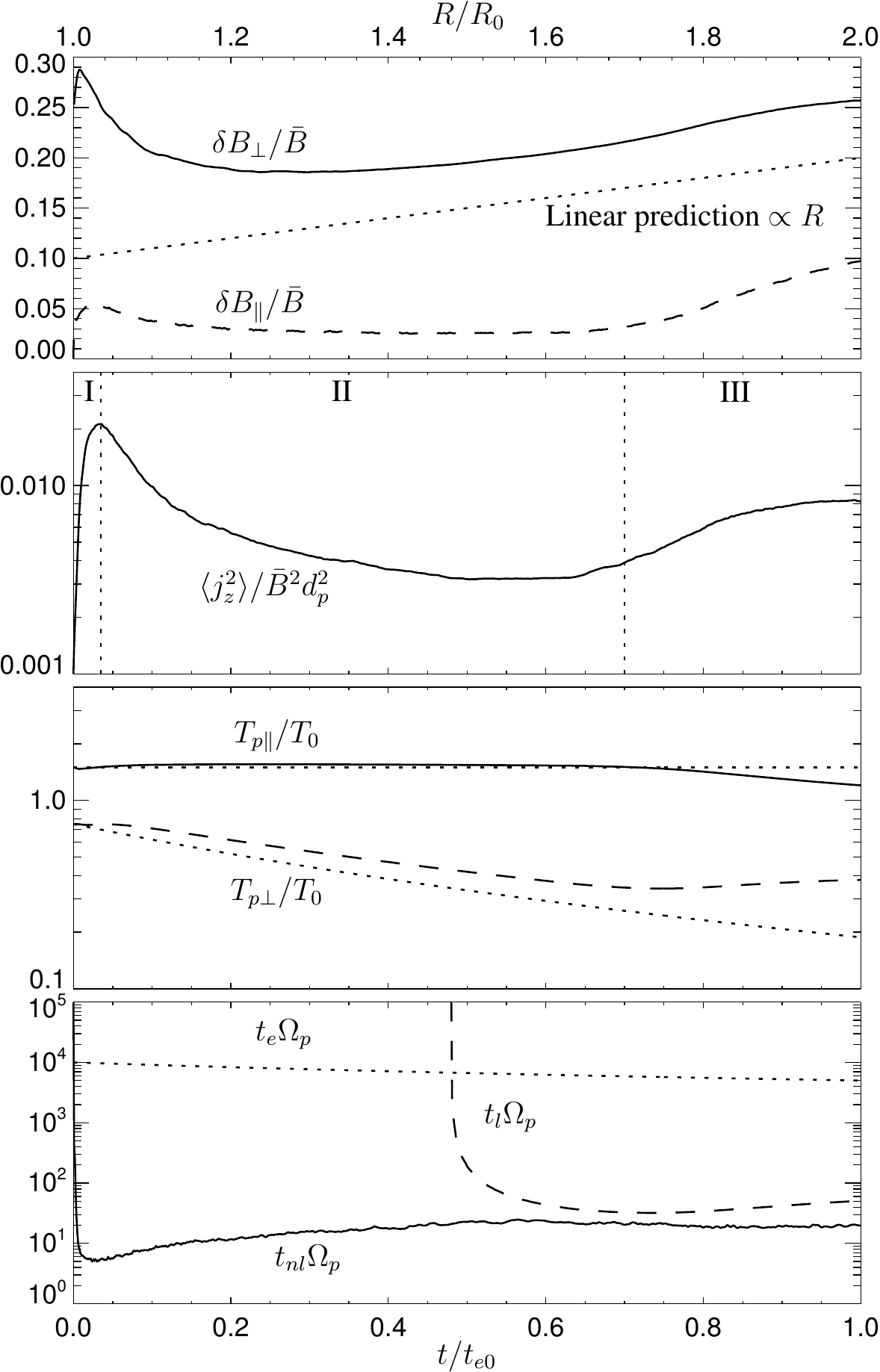}}
\caption{Time evolution of different quantities: (from top to bottom) the fluctuating magnetic field 
(solid) perpendicular $|\delta B_\perp|^2$ and (dashed)  parallel
 $|\delta B_\||^2$ with respect to $\boldsymbol{B}_0$ (the dotted line shows
the linear prediction for the zero-frequency Alfv\'en waves); 
the average squared parallel current $\langle j_z^2\rangle$; 
the parallel $T_{p\|}$ (solid line)
and perpendicular $T_{p\perp}$ (dashed line) proton temperatures 
(the $\|$ and $\perp$ directions are here with respect to the local magnetic
field; the dotted lines denote the corresponding CGL predictions); 
(solid) the nonlinear eddy turnover time $t_{nl}$ at $k d_{p}=1$
(dotted) the expansion time $t_e$, and (dashed) the linear time $t_l$ for the oblique
firehose instability.
\label{evol}}
\end{figure}

Figure~\ref{evol} summarizes the evolution of the simulated system
which goes through three phases. During the first phase the
system relaxes from the initial conditions and turbulence develops; the level of magnetic fluctuations increases
at the expense of proton velocity fluctuations. The fluctuating
magnetic field $\delta B_\perp/\bar{B}$ reaches the maximum at about $t\sim 0.008 t_{e0}$.
During this phase a parallel current $j_z$ 
is generated, $\langle j_z^2 \rangle$ normalized to $\bar{B}^2/d_p^2$ reaches a
maximum at $t\sim0.035 t_{e0}$ indicating a presence
of a well developed turbulent cascade \citep{mipo09,valeal14}.
After that the system is dominated by a decaying
turbulence, the fluctuating magnetic field initially decreases faster
than $\bar{B}$ till about  $0.3 t_{e0}$.

\begin{figure}[htb]
\centerline{\includegraphics[width=8cm]{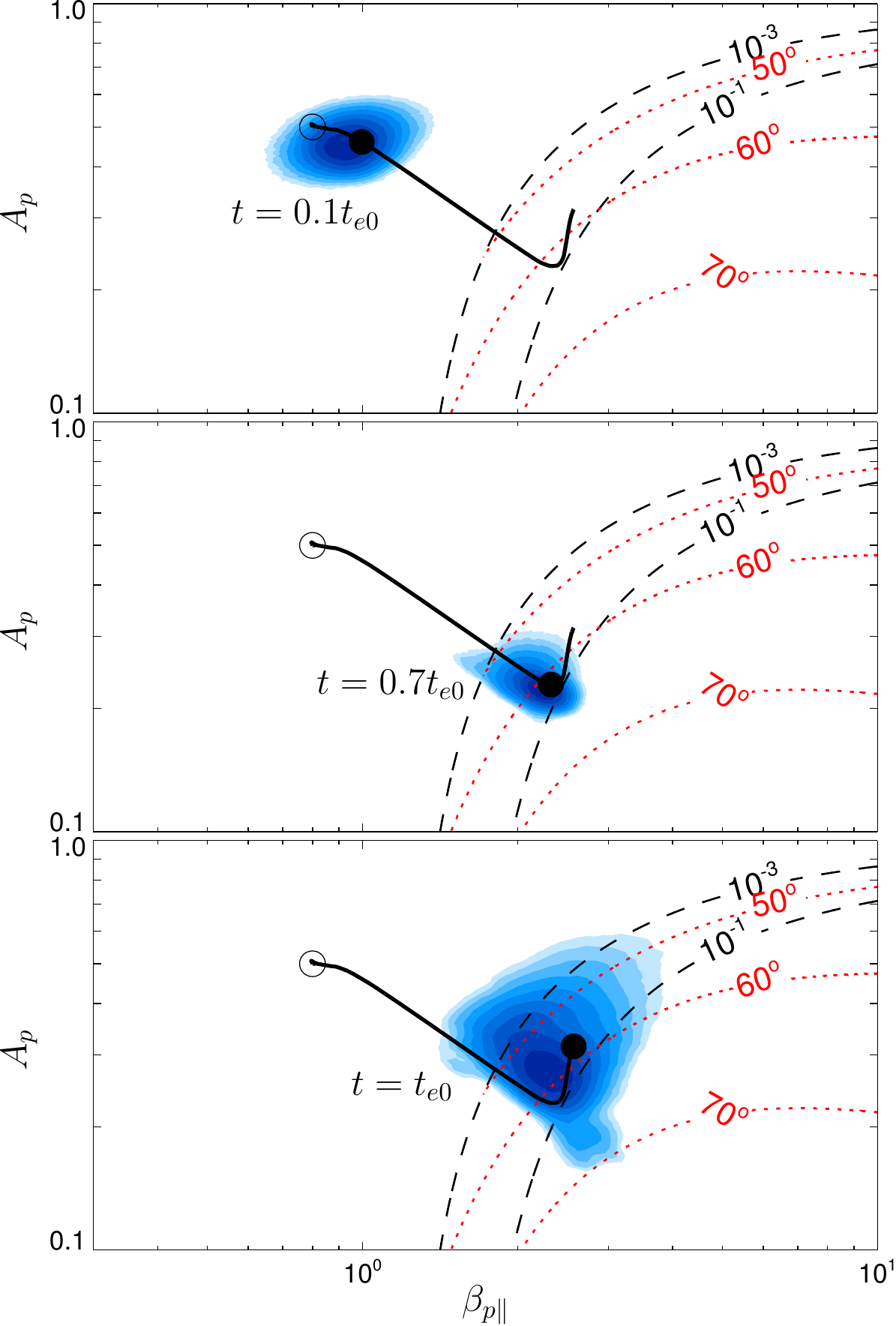}}
\caption{Simulated data distribution in the plane 
$(\beta_{p\|},A_{p})$ at different times.
The empty circles give the initial condition whereas the solid circles
denote the average values. The solid
lines show the evolution of the average values.
The dashed contours show
 the maximum growth rate $\gamma_m$ (in units of $\Omega_{p}$) of the oblique firehose
instability
as a function of $\beta_{p\|}$ and $A_{p}$.
The dotted contours display the corresponding angle of propagation of the most unstable
mode.
\label{bean}}
\end{figure}

During the second phase protons are heated.
For negligible heat fluxes, collisions, and fluctuations, one expects
 the double adiabatic behavior or CGL \citep{chewal56,mattal12}:
the parallel and perpendicular temperatures (with respect to the
magnetic field) are expected to follow $T_{p\perp}\propto \bar{B}$
and $T_{p\|}=\mathrm{const.}$, respectively.
 $T_{p\perp}$ decreases slower than $\bar{B}$ 
during the whole simulation, protons are heated in the perpendicular direction
while in the parallel direction the heating lasts till about $t\sim 0.25 t_{e0}$ whereas
afterwards protons are cooled.
The parallel and perpendicular heating rates could be estimated as \cite[cf.,][]{versal15}:
\begin{equation}
 Q_\|= \frac{\bar{n}^3 k_B}{\bar{B}^2} \frac{\mathrm{d}}{\mathrm{d}t}\left(\frac{T_\| \bar{B}^2}{\bar{n}^2}\right)
 \ \mathrm{and} \  Q_\perp=\bar{n} k_B \bar{B} \frac{\mathrm{d}}{\mathrm{d}t} \left( \frac{T_\perp} { \bar{B}}\right).
\end{equation}
A more detailed analysis indicates that between $t=0.1 t_{e0}$ and $t=0.7 t_{e0}$ the parallel heating rate $Q_\|$
smoothly varies from about $0.2 Q_e$ and $-0.2 Q_e$, whereas 
$Q_\perp$ is about constant $\sim 0.2 Q_e$; here
$Q_e= \bar{n} k_B T / t_e$. In total, protons are heated till $t\sim0.7 t_{e0}$
and the heating reappears near the end of the simulation $t\gtrsim 0.95 t_{e0}$. 
Note that the perpendicular heating rate is
 a nonnegligible fraction of that observed in the solar wind where $Q_\perp \approx 0.6 Q_e$ \citep{hellal13};
however, the proton heating in 2D hybrid simulations is typically quite sensitive to the used
electron equation of state \citep{paraal14} and also to the used resistivity and the number of
particles per cell \citep{franal15b}.
The turbulent heating is, however, not sufficient to overcome the expansion-driven perpendicular cooling
as in the solar wind \citep{mattal07}. 
During the third phase,   $t\gtrsim 0.7 t_{e0}$, 
there is an enhancement of the parallel cooling and perpendicular heating which cannot be ascribed 
to the effect of the turbulent activity. 
For a large parallel proton temperature anisotropy a 
firehose instability is expected. The presence of such an instability is
supported by the fact that the fluctuating magnetic field increases (with respect to
the linear prediction) suggesting a generation of fluctuating magnetic energy at the expense
of protons.
To analyze the role of different processes in the system
we estimate their characteristic times \citep{mattal14b}.
The bottom panel of Figure~\ref{evol} compares 
the turbulent nonlinear 
eddy turnover time 
$t_{nl}=  k^{-3/2} (P_{\boldsymbol{B}\perp}(k)/\mu_0 m_{p})^{-1/2}$ at
$k d_{p}=1$ \cite[cf.,][]{mattal14b}, the expansion time $t_e$, and the 
linear time $t_l$ of the oblique firehose \citep{hema00,hema01} estimated as 
$t_l=1/\gamma_{m}$, where $\gamma_m$ is the maximum growth rate 
calculated
from the average plasma properties in the box assuming bi-Maxwellian
proton velocity distribution functions \citep{hellal06}. 
The expansion time $t_e$ is much longer than 
$t_{nl}$ at $k d_{p}=1$ (as well as at the injection
scales).
The expanding system
becomes theoretically unstable with respect to the oblique firehose
around $t\sim 0.47 t_{e0}$ but clear signatures of a fast proton isotropization
and of a generation of enhanced magnetic fluctuations 
appear later $t\gtrsim 0.7 t_{e0}$. This is about the time when the
linear time becomes comparable to the nonlinear time at ion scales.
After that, $t_{l}\Omega_p$ slightly increases as a result of a saturation of 
the firehose instability whereas
$t_{nl}\Omega_p$ at $k d_p=1$ is about constant (note that $\Omega_p$ decreses
as $R^{-2}$).
This may indicate that the instability has to be fast enough to compete
with turbulence; however, the 2-D system has strong geometrical constraints.
Also the stability is governed by the local plasma properties.
Figure~\ref{bean} shows the evolution of the system
in the plane $(\beta_{p\|},A_{p})$.
During the evolution, a large spread of local values in the
2-D space $(\beta_{p\|},A_{p})$ develops. 
Between $t\simeq 0.1 t_{e0}$ and $t\simeq 0.65 t_{e0}$ the average quantities
evolve in time following 
$\langle A_{p}\rangle \propto \langle \beta_{p\|} \rangle^{-0.86}$.
This anticorrelation is
 qualitatively similar to in situ Helios observations between 0.3 and 1~AU
\citep{mattal07}. During the third stage, when
the strong parallel temperature anisotropy is reduced,
both local and average values of  $\beta_{p\|}$ and $A_{p}$
appear to be bounded by the linear marginal stability conditions 
of the oblique firehose
\citep{hetr08}, although relatively large theoretical growth rates
$\gamma_{m}\sim 0.1 \Omega_{p}$ are expected.

\begin{figure}[htb]
\centerline{\includegraphics[width=8cm]{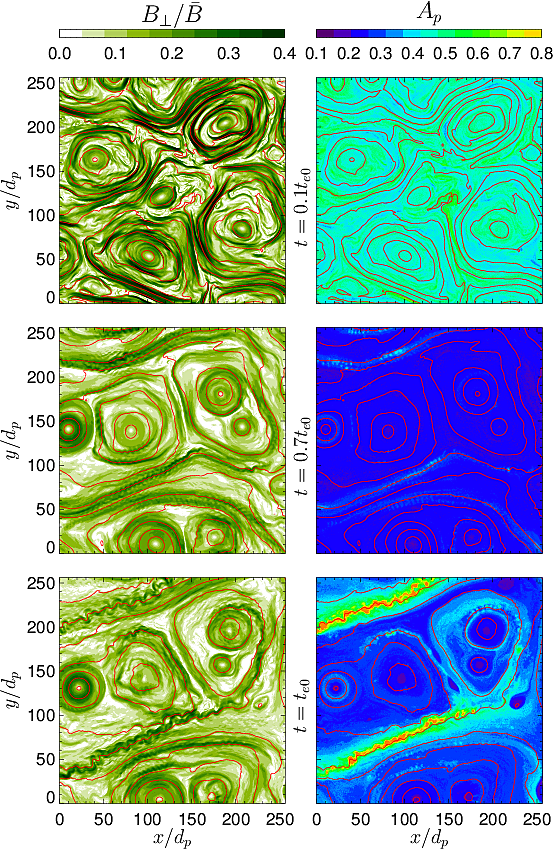}}
\caption{Color scale plots of (left) $\delta B_\perp$ and (right) $A_{p}$
 as functions of $x$ and $y$ for (top) $t=0.1 t_{e0}$,
(middle) $t=0.7 t_{e0}$,  and (bottom) $t=t_{e0}$. The solid lines
show selected (projected) magnetic field lines.
 Only a quarter of the simulation box is shown. 
\label{bban}}
\end{figure}

The evolution in the real space is shown in 
Figure~\ref{bban} which
 shows the magnitude of the perpendicular fluctuating magnetic field $\delta B_\perp$
and the proton temperature anisotropy $A_{p}$ at different times
(see also the movie which combines the evolution in Figures~\ref{bean} and~\ref{bban}). 
The modes with initially random phases
rapidly form vortices and current sheets. Despite the overall turbulent heating
a strong temperature anisotropy $T_{p\perp}<T_{p\|}$ develops owing to the expansion and
a firehose-like activity develops in a form of localized waves/filaments
with enhanced $\delta B_\perp$. 
These fluctuations appear in regions between vortices where $B_\perp/\bar{B}$
is enhanced, i.e., in places where the angle between
the simulation plane and the local magnetic field $\theta_B$ ($\approx \arccos B_\perp/\bar{B}$) is less oblique
(reaching $\sim 60^\mathrm{o}$ and below). This is in agreement with the theoretical expectations,
while the oblique firehose is unstable for moderately oblique wave vectors with respect to the magnetic field 
near threshold, further away from threshold the unstable modes 
become more oblique (see Figure~\ref{bean}) and these oblique angles are locally available between vortices where we observe
the enhanced level of magnetic fluctuations. These geometrical factors may be responsible for the late appearance
of the instability (but constraints on the instability time scales imposed by the turbulent non-linearities are likely also important).  The localized wave packets are Alv\'enic,
have wavelengths of the order of $10 d_p$ and propagate with a phase velocity
about $0.1 v_A$ in agreement with expectation for the nonlinear phase of the oblique firehose.
Furthermore, the parallel temperature anisotropy is strongly reduced in their vicinity.
These Alv\'enic wave packets are responsible for the enhanced level of the magnetic PSD at ion scales seen
in Figure~\ref{spec}.

For a linear instability it is expected that the magnetic fluctuations increase
exponentially in time during its initial phase \cite[except when
the growth time is comparable to the expansion time, cf.,][]{teve13}.
Figure~\ref{evol} however shows that the overall magnetic fluctuations $\delta B_\perp /\bar{B}$ 
(with respect to the linear prediction) and $\delta B_\| /\bar{B}$
increase rather slowly (secularly) in time for $t\gtrsim 0.7 t_e$. This behavior is expected for a long time evolution
in a forced system after the saturation \cite[cf.,][]{mattal06,rosial11,kunzal14}.
An additional analysis indicates that the expected exponential growth is indeed seen in the simulation
but only locally both in space and time.
This exponential growth is obscured by the turbulent fluctuations and, furthermore, it
is blurred out due to the averaging over the simulation box in the global view of Figure~\ref{evol}.

\begin{figure}[htb]
\centerline{\includegraphics[width=8cm]{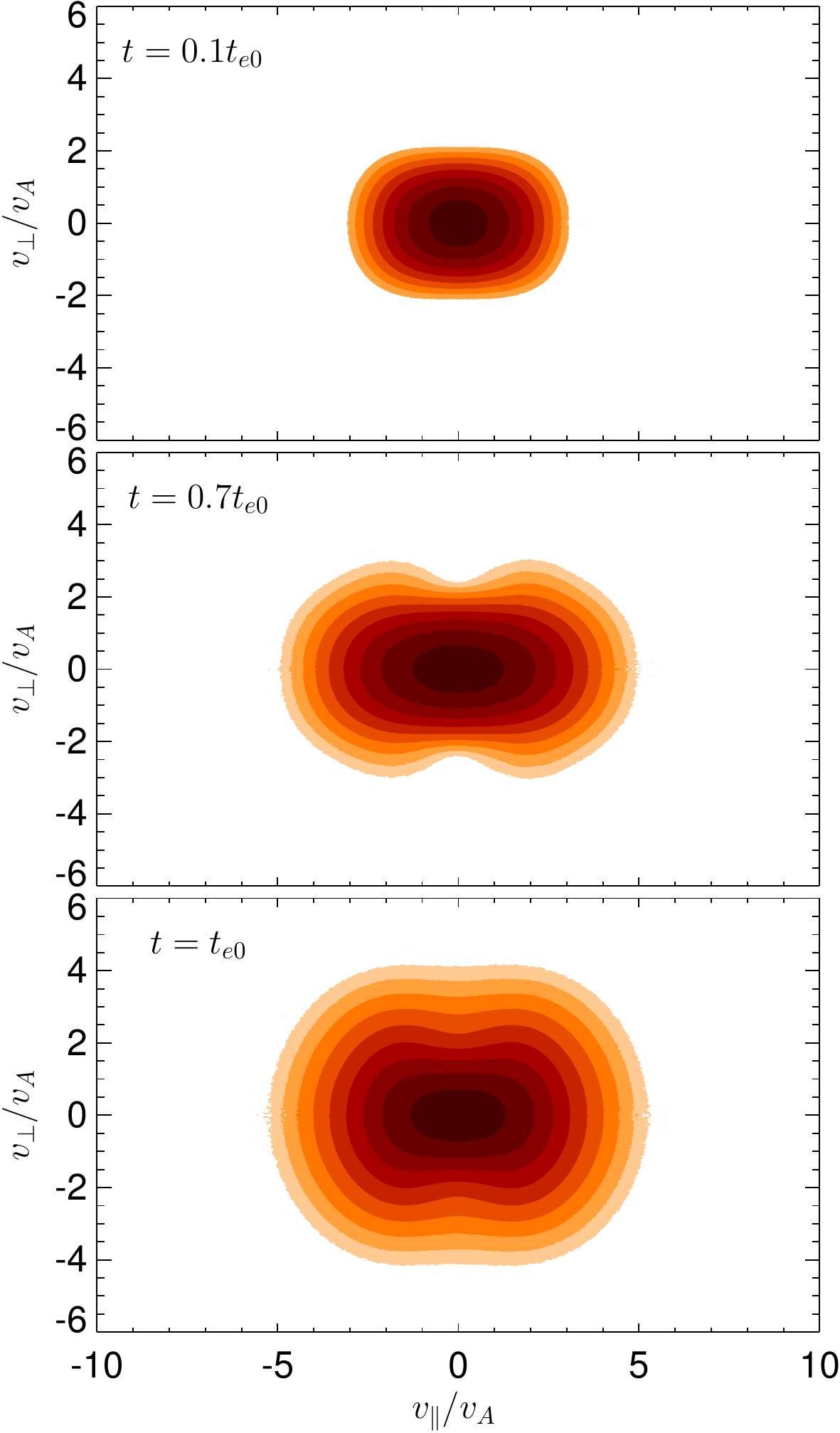}}
\caption{Average proton velocity distribution function $f$
as a function of parallel and perpendicular velocities $v_\|$ and $v_\perp$
(with respect to the local magnetic field) for for (top) $t=0.1 t_{e0}$,
(middle) $t=0.7 t_{e0}$,  and (bottom) $t=t_{e0}$.
\label{vdf}}
\end{figure}

On the microscopic level the firehose activity leads to 
an efficient scattering from parallel to perpendicular direction 
of protons in the velocity space. Figure~\ref{vdf} shows the evolution
of the proton velocity distribution function $f=f(v_\|,v_\perp)$ averaged
over the simulation box.
While turbulence leads locally to a complex proton distribution functions 
\cite[cf.,]{valeal14,serval15} the average proton distribution 
function during the first two phases remain relatively close
to a bi-Maxwellian shape (Figure~\ref{vdf}, top panel).
During the third phase there appear
clear signatures of the cyclotron diffusion
(for protons with $v_\| \gtrsim v_A$) as expected for the oblique firehose 
instability \citep{hetr08}.

\section{Discussion}
Using 2-D hybrid simulations we investigated
the evolution of turbulence in a slowly expanding plasma. 
The numerical model shows that the turbulent heating
is not sufficient to overcome the expansion driven cooling  and 
that the oblique firehose becomes active for a sufficiently large parallel proton temperature
anisotropy and for sufficiently oblique angles of propagation.

While the modeled expansion  is about ten times faster than 
in the solar wind, the ratio between the expansion and
the nonlinear eddy turnover
  time scales is quite realistic: $t_e/t_{nl}\approx 1000$ at $kd_p=1$ for $t\gtrsim 0.7 t_e$
which is about four times smaller than that of the solar
wind with similar plasma parameters at 1~AU \citep{mattal14b}.
Note also that a similar evolution is
observed for many different plasma and expansion parameters.

In the present case, both turbulence and the 2-D geometry constraints strongly affect 
the firehose instability and there are indications that firehose has
an influence on turbulence (the mixed third-order structure functions \citep{verdal15}
are enhanced due to the firehose activity suggesting a stronger cascade rate).  
The problem of the interaction between turbulence and kinetic instabilities
requires further work;  
three-dimensional simulations are needed to
investigate the interplay between turbulence and instabilities
as usually the most unstable modes are parallel or moderately oblique with respect to the ambient magnetic field;
in the present case the parallel firehose \citep{garyal98,mattal06} would be the dominant instability  but the
2-D constraints strongly inhibit it. On the other hand, numerical simulations indicate that
the oblique firehose plays an important role in constraining the proton temperature
anisotropy in the expanding solar wind even in the case when the parallel firehose is dominant \citep{hetr08}.
Nevertheless,
the present work for the first time clearly demonstrates 
that kinetic instabilities may coexist
with strong plasma turbulence and bound the plasma parameter space.

\acknowledgments
The authors wish to acknowledge valuable discussions
with Marco Velli.
The research leading to these results has received funding from the
European Commission's 7\textsuperscript{th} Framework Programme under
 the grant agreement \#284515 (project-shock.eu).
PH and PMT acknowledge GACR grant 15-10057S
and the projects RVO:67985815 and RVO:68378289.
LM acknowledges UK STFC grant ST/K001051/1. 
AV acknowledges the Interuniversity Attraction Poles Programme initiated 
by the Belgian Science Policy Office (IAP P7/08 CHARM).

\end{document}